\documentclass[paper]{JHEP}
\usepackage{epsf}
\usepackage{amsmath}
\usepackage{amsfonts}

\newcommand{\pa}{\partial}
\newcommand{\tr}{{\rm tr}}
\newcommand{\comment}[1]{}

\newcommand{\CO}{{\cal O}}
\newcommand{\pasl}{\pa\kern-.55em /}

\newcommand{\ksl}{k\kern-.55em /}
\newcommand{\twfld}{\sigma_6\sigma_7\sigma_8\sigma_9 e^{ik\cdot X}}

\DeclareFixedFont{\xiiss}{OT1}{cmss}{m}{n}{12}
\DeclareFixedFont{\ixss}{OT1}{cmss}{m}{n}{9}
\DeclareFixedFont{\cmrnine}{OT1}{cmr}{m}{n}{9}
\newcommand{\field}[1]{\mathbb{#1}}
\newcommand{\CC}{{\field C}}

\newcommand{\ZZ}{{\field Z}}
\newcommand{\QQ}{{\field Q}}
\newcommand{\CCs}{\hbox{\ixss C\kern-.4emI}}
\newcommand{\ZZs}{\hbox{\ixss Z\kern-.4emZ}}
\newcommand{\tors}{\eta}

\newcommand{\littlefig}[2]{
	\epsfxsize=#2in
	\epsfbox{#1}
}

\keywords{Discrete Torsion,Duality in Gauge Field Theories,D-branes,Brane Dynamics in Gauge Theories}
\preprint{ILL-(TH)-00-01\\ hep-th/0001055}
\title{Discrete Torsion, AdS/CFT and Duality}
\author{David Berenstein\thanks{\email{berenste@pobox.hep.uiuc.edu}}
and Robert G. Leigh\thanks{\email{rgleigh@uiuc.edu}}\\
Department of Physics\\
University of Illinois at Urbana-Champaign\\
Urbana, IL 61801}

\abstract{We analyse D-branes on orbifolds with discrete torsion,
extending earlier results. We analyze certain Abelian orbifolds  of the
type $\CC^3/\Gamma$, where $\Gamma$ is given by $\ZZ_m\times \ZZ_n$, for
the most general choice of discrete torsion parameter. By comparing with
the AdS/CFT correspondence, we can consider different geometries which
give rise to the same physics.  This identifies new mirror pairs and
suggests new dualities at large $N$. As a by-product we also get a more
geometric picture of discrete torsion.}

\begin{document}

\section{Introduction}

Orbifold singularities \cite{DHVW,DHVW2} have been the object of much
attention in string theory, as they provide solvable examples of
conformal field theories and correspond to degenerate limits of
Calabi-Yau families of vacua. Early on it was noticed that on
calculating the spectrum of an orbifold there is an inherent ambiguity
in the phases chosen for the twisted sector.\cite{V} This ambiguity,
discrete torsion, is classified by the group $H^2(\Gamma,U(1))$, where
$\Gamma$ is the orbifold group.

Discrete torsion has been analyzed in the context of deformation theory
in \cite{VW,AMG} where it was noticed that discrete torsion
singularities cannot be geometrically resolved, and that they provide totally
distinct branches for compactifications of string theory. As a
parameter, discrete torsion can be naturally associated to the $NSNS$
2-form in string theory, and thus it is natural to interpret it in terms
of gerbes.\cite{S}

More recently, it has been shown\cite{D,DF} that discrete torsion can be
analyzed in terms of D-branes at the orbifold by taking into account
projective representations of the orbifold group. This is also needed
for Matrix theory setups, as in \cite{HW}. In this way one builds quiver
theories \cite{DM} for the modified orbifold which have a totally
different content than the theories without discrete torsion. This
analysis has been extended to the boundary state formalism in \cite{DG}.

A new tool which to date has not been exploited to analyze this
situation is to make use of the AdS/CFT correspondence \cite{M}. In
particular, the large $N$ limit of D3 branes at an orbifold was
discussed in \cite{KS}, but discrete torsion not considered. In this
paper we analyze orbifolds with discrete torsion within the AdS/CFT
correspondence.

The first task is to describe the physics of a particular type of
orbifold with $\Gamma= \ZZ_m\times \ZZ_n$. We find the low energy field
theory decsription of D3-branes at the singularity with discrete torsion
for all possible choices of discrete torsion. This is done in Sections
\ref{sec:orbifold},\ref{sec:dbranes}. The theories obtained are
supersymmetric four-dimensional field theories of the brane-box type,
with a modified superpotential. The orbifold can be recovered by the
moduli space of vacua, which depends on the superpotential. This
calculation is done in Section \ref{sec:moduli}. We finish the field
theory study in Section \ref{sec:spectrum} where we compute the spectrum
of chiral primaries of the CFT at the origin.

Secondly, in Section \ref{sec:ADS/CFT} we study the large $N$ theories
associated to the orbifold with discrete torsion. The first fact that we
need to consider is that they lie in the moduli space of deformations of
more standard quiver diagrams, as can be argued along the lines of
\cite{LS}. By following the deformations in supergravity, we are able to
find a duality of type IIB string theories at weak coupling compactified
on very different spaces. This is an example of mirror symmetry for RR
backgrounds.

Subgroups of $\Gamma$ leave 2-planes fixed, which descend to fixed
circles in the large $N$ limit. By a careful analysis of the string
theory at the orbifold one can find the correct boundary conditions for
the corresponding twisted sector states. These states are massless and
hence survive in the supergravity limit. Thus discrete torsion is
encoded in the near-horizon geometry, through different choices of
boundary conditions for twisted fields. One can compare these results
with the field theory from Section \ref{sec:spectrum}, and agreement is
found. As a by-product we obtain geometric information on the orbifold
with discrete torsion.

The analysis shows that if one considers resolutions by blowups of the
classical geometry, on going around the fixed circles they are affected
by monodromy. The monodromy encodes the discrete torsion, and shows that
the topology of the $S^5/\Gamma$ is very different for the cases with
and without discrete torsion. In particular, with a non-trivial
monodromy one finds torsion classes in the homology, and then the
discrete torsion parameter is the B-field flux around this homology
two-cycle. This difference in topology  also explains why the
deformation theory of the orbifolds with and without discrete torsion
are so different.

Section \ref{sec:Dualities} deals with new dualities for large $N$ field
theories. Two theories with the same field content at different values
of marginal couplings have the same moduli space of flat directions and the same
spectrum of chiral primary operators, and, furthermore, the coupling
that distinguishes them does not appear in planar diagrams. 
The duality acts
discontinuously on the space of couplings by a permutation of the roots
of unity.

\section{The orbifold}\label{sec:orbifold}

We  analyze orbifolds of the Type IIB superstring
of the form  $\CC^3/\Gamma$ with $\Gamma
= \ZZ_m\times \ZZ_n$, and whose generators act on the coordinates by
\begin{eqnarray}
e_1:(z_1, z_2, z_3) \to (\alpha z_1, \alpha^{-1} z_2, z_3)\\
e_2:(z_1, z_2, z_3) \to (z_1, \beta z_2, \beta^{-1} z_3)
\end{eqnarray}
with\footnote{Throughout, we use the notation $\omega_k\equiv e^{2\pi
i/k}$.} $\alpha=\omega_m$  and $\beta = \omega_n$.

Because of the choice of action of the orbifold, we
preserve $N=2$ supersymmetry in four dimensions, which can  be broken
to $N=1$ by the addition of D-branes at the singularity.
 
The discrete torsion parameter, is a `bilinear' 
element  $\epsilon \in 
H_2(\Gamma, U(1))$ such that
\begin{eqnarray}
\epsilon(g_1, g_2 g_3) &=& \epsilon(g_1, g_2) \epsilon(g_1, g_3)\\
\epsilon(g_1, g_2) &=& \bar\epsilon(g_2, g_1)\\
\epsilon(g_1,g_1)&=&1
\end{eqnarray}
for $g_1, g_2, g_3 \in \Gamma$. 
As $\Gamma$ has two generators, $\epsilon$ is completely determined by
the number 
\begin{equation}
\tors = \epsilon(e_1, e_2)
\end{equation}

It is easy to see that the solution of these equations is as follows. If
$m,n$ are relatively prime, there is no discrete torsion. If we denote
$p=gcd(m,n)$, then we find $H_2(\Gamma, U(1))=\ZZ_p$. This includes the
oft-studied special case $m=n$, where clearly $p=n$. The discrete
torsion is thus determined by $\tors=\omega_p^r$, where
$r=0,1,\ldots,p-1$. Let $s$ be the smallest non-zero number such that
$\tors^s = 1$. It is the integer $s$ which determines the physics, and
for each value of $s$ the orbifolds will behave differently; for
example, they will have a different resolution of singularities.

For later use, it will be useful to define a common root of unity
$\lambda$, such that $\lambda^{mn}=1$. This phase will appear in the
superpotential of the low energy gauge theory. It will also be useful to
define $q=lcm(m,n)$, the least common multiple of $m,n$.

\section{D-branes on the orbifold}\label{sec:dbranes}

We want to locate D3-branes at the orbifold. The construction for the
low energy effective action follows \cite{DM, D}. In order to
incorporate the discrete torsion parameter into the field theory, one
has to consider all possible projective representations of the orbifold
group with the same cocycle. That is, one considers representations
where 
\begin{equation} \label{eq:disctors}
\gamma(e_1) \gamma(e_2) = \tors \gamma(e_1e_2)
\end{equation}

It is easy to see that given a projective representation $R$ of the
group $\Gamma$ and a non-projective representation $\chi$, the
representation $R\otimes \chi$ is a projective representation of
$\Gamma$ with the same discrete torsion parameter as $R$. By tensoring
$R$ with all possible representations of $\Gamma$ we can obtain all
possible projective representations of $\Gamma$. 

We can thus build the analog of the regular representation of $\Gamma$
by considering $R= \oplus_i \dim (R_i)R_i$, with $R_i$ all the unitarily
inequivalent projective representations of $\Gamma$, and out of the
regular representation we should obtain the quiver diagram representing
a D-brane which is outside the singularity. Any other D-brane
configuration can be obtained by choosing the representation to be
non-regular \cite{DDG,BCD}. String consistency imposes anomaly
cancellation on the configuration.

One irreducible projective representation of $\Gamma$ is given 
in terms of $s\times s$ matrices
\begin{equation}
\gamma(e_1) = 
\hbox{diag}( 1, \tors^{-2}, \tors^{-4}, \cdots, \tors^{-2(s-1)}),
\quad \gamma(e_2) = \begin{pmatrix} 0&1&0&\cdots&0\\
0&0&1&\dots&0\\
\vdots&\vdots&\vdots&\ddots&\vdots\\
1&0 &0&\cdots& 0
\end{pmatrix}
\end{equation}
Any other irreducible representation is given by multiplying
$\gamma(e_1)$ by an $m^{th}$ root of unity $\alpha^k$ and $\gamma(e_2)$
by an $n^{th}$ root of unity $\beta^\ell$. This is equivalent to
multiplying the representation $R_i$ by a character of the orbifold
group, and thus producing another projective representation of the
group, which is also irreducible.
Two representations labeled by $(k,\ell)$ and $(k',\ell')$ are
equivalent if $k= k' \mod m/s$ and $\ell = \ell' \mod n/s$, as can be
seen by decomposing the representation into either characters of 
$\ZZ_m$ or $\ZZ_n$.

As all the irreducible projective representations are $s$-dimensional,
we obtain the regular representation by taking all possible different
projective representations with the same cocycle, multiplied by the
dimension of the representation.
\begin{equation}
R= \oplus_{k< m/s,\ \ell<n/s}\  s R_{k\ell}
\end{equation}
The total dimension of $R$ is $mn$, which is the order of the group,
$|\Gamma|$.

The gauge theory living on the
D-branes is given by a quiver diagram with $\frac{m}{s}\times\frac{n}{s}$
nodes, one for each representation. 
This theory is dual to an elliptic brane-box, as in
\cite{HZ,HU}, a part of which appears in Figure 1.

\FIGURE{\littlefig{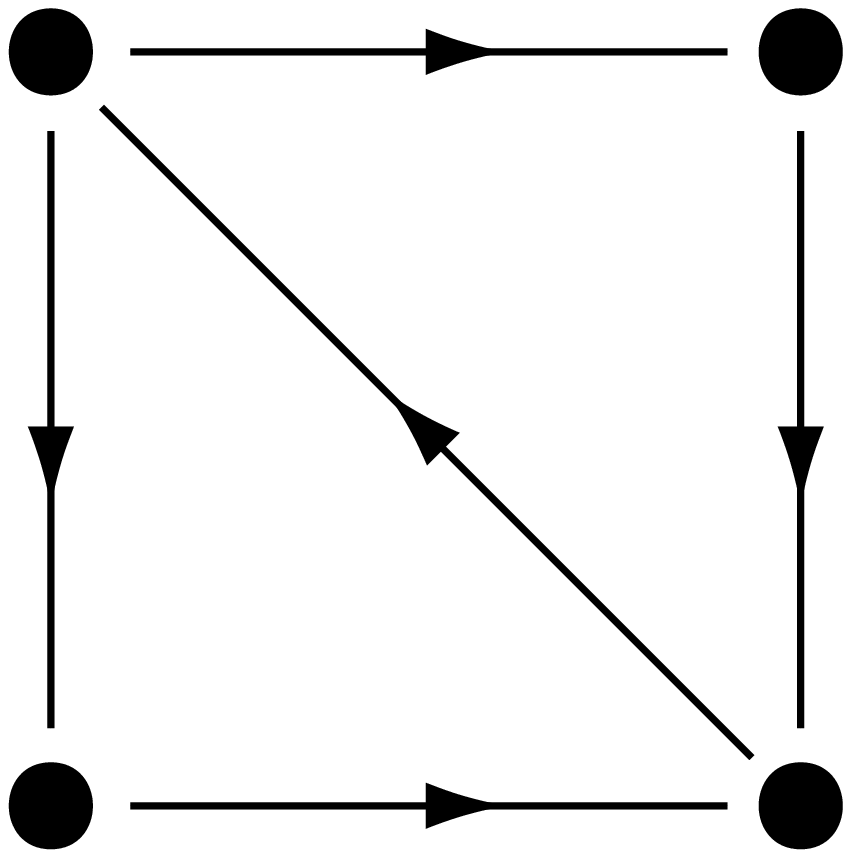}{1.5}
\caption{Part of the brane-box quiver.}}

The nodes indicate the different $(k,\ell)$ gauge groups
corresponding to each irreducible representation. To each of them we
associate the gauge group $U(N_{k\ell})$. For the regular
representation, all of the gauge groups have equal rank, in particular
$N_{k\ell}=s$. The arrows represent chiral multiplets transforming in
the $(N, \bar N)$ of the groups it connects. The
nodes are connected by multiplications by specific characters
of $\Gamma$. 

Let us refer to the chiral multiplets corresponding to 
the horizontal lines as $\phi_{1k\ell}$, the vertical ones
as $\phi_{3k\ell}$ and the diagonal ones as $\phi_{2k\ell}$.
To fully specify the theory, one also needs to give the superpotential.
As brane box diagrams dual to theories with discrete torsion and without
discrete torsion coincide, they can only differ in the superpotential
\cite{D}.

The superpotential is of the form 
\begin{equation}\label{eq:superpot}
\sqrt 2 g \tr (\phi_1\phi_2\phi_3 - \lambda^s \phi_1\phi_3\phi_2)
\end{equation}
A cubic term appears for each small triangle in the quiver, with a
phase difference, depending on the orientation of the given triangle.
Notice that it is the phase $\lambda$ that appears in the
 superpotential, 
instead of $\tors$. To see that this is correct, consider rephasing
all of the chiral fields by arbitrary phases. This transformation would
induce superpotential terms with a set of couplings $\lambda_{C}$ 
($\lambda_{A}$) for (anti-)clockwise triangles. 
\begin{equation}
\sqrt 2 g \tr (\lambda_A\phi_1\phi_2\phi_3 - \lambda_C \phi_1\phi_3\phi_2)
\end{equation}
There is one combination
of couplings which is invariant under field redefinitions, and hence
cannot be eliminated. It is the quotient
\begin{equation}\label{eq:quotient}
 \frac{\prod_{\Delta_C}\lambda_C}{\prod_{\Delta_A}\lambda_A}=\tors
\end{equation}
and turns out to equal the discrete torsion. In this way, the discrete
torsion is encoded in the theory. With our choice of conventions,
$\lambda_C=\lambda^s, \lambda_A=1$. We have chosen the couplings so that
they all are the same, and therefore the quantum symmetry is easier to
construct. We could just as well have made all the couplings equal
except at one triangle, where we would insert the phase $\tors$. Notice
that when $m=n$ with only one node in the diagram, this reduces to the
case considered in \cite{D,DF}.
\FIGURE{\littlefig{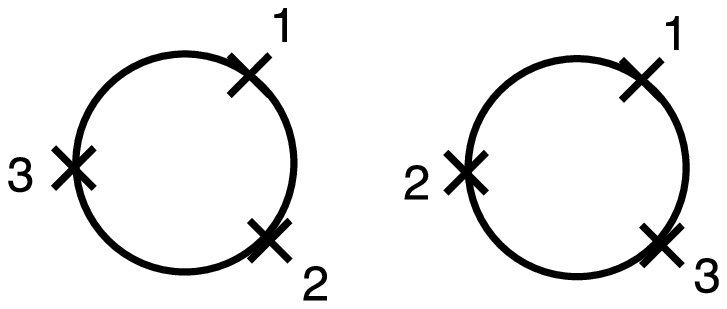}{2}
\caption{The two disc diagrams.}\label{fig:discs}}

It should be possible to deduce the form of the superpotential directly
through string computations. In particular, we wish to show that the diagrams
in Figure \ref{fig:discs} differ essentially in phase.

In the field theory, there is a
vacuum diagram which encodes the phase $\eta$, and can thus be traced back
to the superpotential (\ref{eq:superpot}). It uses each of the cubic vertices
exactly once, and thus is proportional, in field theory, to the ratio
of phases (\ref{eq:quotient}) (or its inverse, depending on chirality). 
When drawn on top of the quiver, it has
the topology of a net of hexagons (one for each node), as shown in Figure 
\ref{fig:hexagon}.
\FIGURE{\littlefig{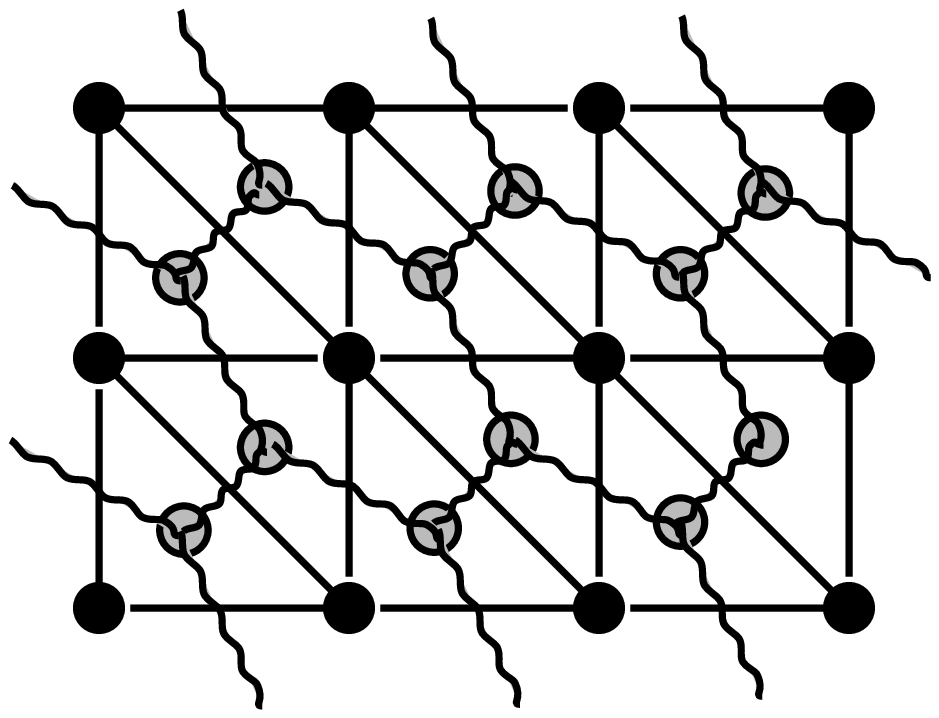}{2}
\caption{A portion of the dual hexagon Feynman diagram.}\label{fig:hexagon}}

In string theory, we would evaluate this using projected fields
$\phi_{a,inv}=\sum_{g\in\Gamma}r_a(g)\gamma^{-1}(g)\phi_a\gamma(g)$. The
diagram is equivalent to a genus one worldsheet with $mn/s^2$ holes (one for
each node in the quiver). Thus the result is a product of traces, one for
each node labelled $i,j$, each of
the form
\begin{equation}
\tr_{R_{ij}}\ \gamma^{-1}(g_{ij,1})\gamma(g_{ij,2})\gamma^{-1}(g_{ij,3})\gamma(g_{ij,4})
\gamma^{-1}(g_{ij,5})\gamma(g_{ij,6})
\end{equation}
where the group elements are paired, viz $g_{i,j,3}=g^{-1}_{i+1,j,6},\ 
g_{i,j,4}=g^{-1}_{i+1,j+1,1},\ g_{i,j,5}=g^{-1}_{i,j+1,2}$.
This result is the analogue of the single node computation of Ref. \cite{D}.

It should be possible to reduce the product of traces by implementing the
relation (\ref{eq:disctors}) and using character formulas. 
Furthermore, the results given later in the paper
are fully consistent with the form of the superpotential, (\ref{eq:superpot}).

\section{The moduli space of a regular D-brane}\label{sec:moduli}

In order to verify that we have the correct field theory, we should
consider the moduli space of vacua for a regular D-brane. 
In particular, one should have a three dimensional
component where the D-brane lives away from the singularities, which
corresponds to the total orbifold space  minus the fixed lines. 
When the D-brane approaches one of the fixed lines, or the fixed point,
the D-brane can fractionate, and the dimension of the moduli space jumps.

{}From the orbifold point of view, the moduli space is described by six
variables with three relations. Only the $\Gamma$-invariant polynomials
in the variables are good functions on the space. This polynomial ring
is generated by $x= z_1^m$, $y=z_3^n$, $z= z_2^{q}$, $u=(z_1z_2)^n$,
$v=(z_2z_3)^m$ and $w=z_1 z_2 z_3$, where $q$ is the least common
multiple of $m,n$. There are three relations\footnote{Note that when
$m=n=q$, the variables $u,v$ are redundant, and the system reduces to
just $f_1=0$.}
\begin{eqnarray}\label{eq:algebraica}
f_1&=&x^{q/m} y^{q/n} z - w^{q}=0\\
f_2&=& vx-w^m=0\label{eq:algebraicb}\\
f_3&=& uy-w^n=0\label{eq:algebraicc}
\end{eqnarray}
and we have three complex lines of singularities where two out of the
three $x,y,z$ are equal to zero. These are the lines which are fixed by
some elements of the orbifold.

{}From the gauge theory, we need to construct the equivalent of
variables $x$, $y$, $z$, $u$, $v$, $w$ in terms of gauge invariant
polynomials of the chiral fields. As is well known from Matrix theory
\cite{BFSS}, one should hope to replace the equations
(\ref{eq:algebraica})--(\ref{eq:algebraicc}) by matrix versions. If the
matrices are made to commute, then we can identify variables
$x,y,z,u,v,w$ for individual branes by their eigenvalues. In particular,
if we want branes in the bulk, we should require that $x,y,z$ be all
invertible (in this case as matrices).

Referring to the quiver, the natural gauge invariant coordinates are
given by traces of closed loops in the diagram. Matrix coordinates will
be given by taking the same  closed loops without the trace, as they
represent matrices whose eigenvalues are gauge invariant quantities,
since gauge transformations act by similarity transformations.

Let us consider the variables $\tilde x_j= \prod^{m/s}_{i=1}\phi_{1ij}$,
and $\tilde y_j=\prod^{n/s}_{i=1}\phi_{3ji}$, and similarly $\tilde z_j$
along diagonals. It is easy to see that all the $\tilde x_i$ are
isospectral up to phases, the same being true for all the $\tilde y_j$,
under the assumption that all the chiral matrices that appear in the
diagram are invertible and satisfy the $F$-constraints. The proof goes
by making use of moves which relate different matrices, and by
application of the Cayley-Hamilton theorem. An example of a move is
shown in Figure \ref{fig:moves}.

\FIGURE{\littlefig{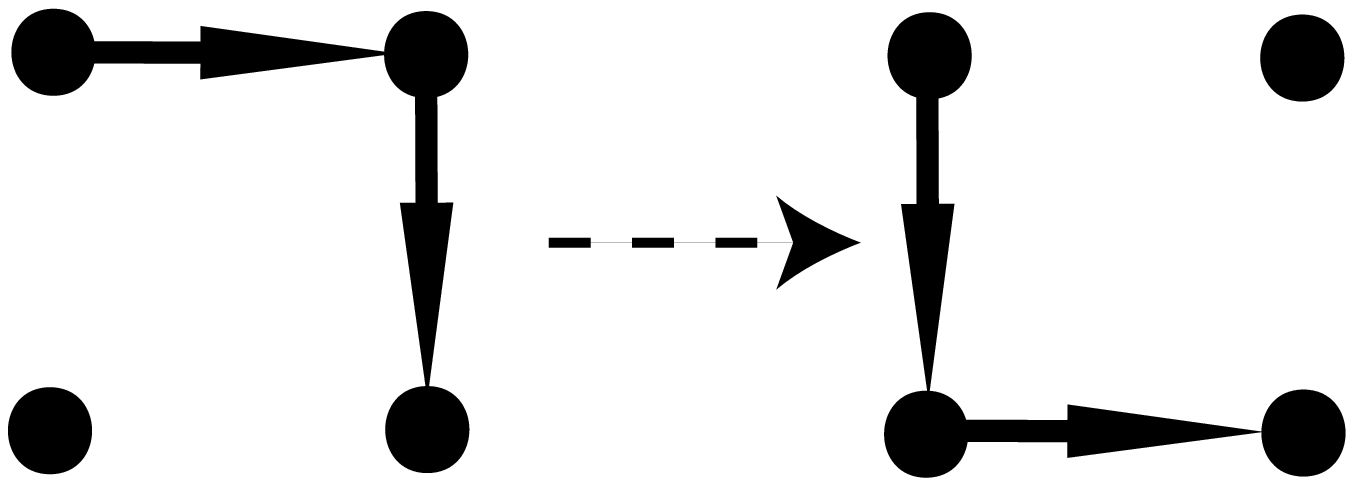}{2}
\caption{Allowed moves}\label{fig:moves}}

The simple moves are given by solving the $F$ constraints for a single
chiral multiplet. Under such a move, we get a phase given by the ratio
of the two associated triangle couplings in the superpotential.

The eigenvalues of the matrix $\tilde x_j$ are determined by the roots
of the characteristic polynomial of $\tilde x_j$. As $\tilde x_j$
satisfies its own characteristic polynomial, we have an identity
$P(\tilde x_j)=0$, which we can use as a matrix equation. We can now
change a row by multiplying by $\phi_2$ at the end. As we want a matrix
equation that begins and ends on the same node, we can complete it by
multiplication by the rest of the triangle that returns the operator to
the starting node. We can now make use of the moves to shift the
polynomial equation one step down, as shown in the Figure
\ref{fig:step}.

\FIGURE{\littlefig{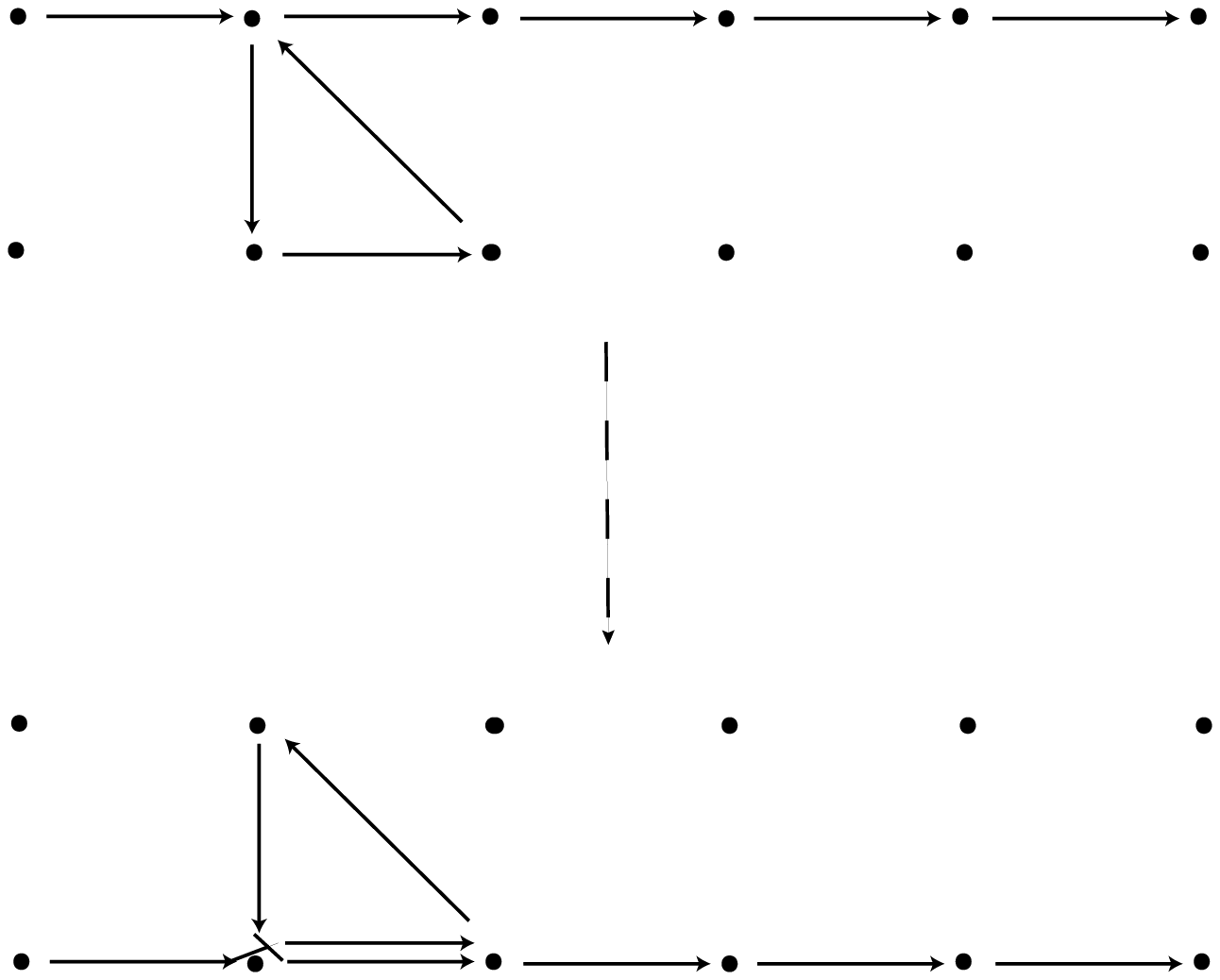}{2}
\caption{Relating the rows.}\label{fig:step}}

Indeed, apart from a global phase in the roots, the two characteristic
polynomials are the same, and therefore the two matrices have the same
content. If we multiply by the triangle operator and use the $F$-term
constraints, we can move the polynomial equation along the rows, and
hence all of the $\tilde x_j$ have the same roots (modulo phases), and
therefore they are equivalent when considered as matrix variables.
Similarly for all the $\tilde y_k$ and $\tilde z_k$. As a result, we
need only consider the matrices $\tilde x_1$, $\tilde y_1$, $\tilde
z_1$.

Now, using the moves, we can also see using (\ref{eq:quotient}) that 
\begin{equation}\label{eq:commutation}
\tilde x_1 \tilde y_1 = \tors \tilde y_1 \tilde x_1
\end{equation}
This means that if $\tilde y_1$ is invertible, then the eigenvalues of
$\tilde x_1$ are only well defined modulo $\tors$, and hence the trace
of $\tilde x_1$ vanishes, so $\tilde x_1$ alone is not a good matrix
variable to describe the moduli space. On the other hand, it shows us
also that the matrices must be of order $s\times s$ in order to find a
representation of the algebraic equation (\ref{eq:commutation}). This is
exactly the size of the matrices in the regular representation of the
quiver.

Now we can take $x= \tilde x_1^s$, and similarly $y=\tilde y_1^s$ and $z
= \tilde z_1^s$. These matrices will commute amongst themselves, and
they have $s$ identical eigenvalues. The matrix $w$ is constructed from
any small triangle in the graph, and it is easy to see that $w$ commutes
with all $x,y,z$. $w$ is defined up to a phase depending only on the
orientation of the triangle.

One can then see that the matrices $x,y,z,w$ satisfy eq.
(\ref{eq:algebraica}) by making use of the $F$-constraint moves. As the
matrices $x,y,z,w$ are all $s\times s$  and proportional to the
identity, we can identify the eigenvalues with the coordinates $x,y,z,w$
of the algebraic variety which describes the orbifold. The two different
choices of $w$ differ by a root of unity which cancels in the equation,
as the order of $\lambda^s$ divides $q$. Similar analyses can be used to
construct the constraints (\ref{eq:algebraicb})--(\ref{eq:algebraicc}),
using the zigzags $u_j=\prod_{i=1}^{n/s} \left(
(\phi_1\phi_2)_{ij}\right)^s$ and $v_i=\prod_{j=1}^{m/s} \left(
(\phi_2\phi_3)_{ij}\right)^s$.

For a total number of branes which is $N$ times that of the regular
representation, we get mutually diagonalizable matrices whose
eigenvalues have a common multiplicity of $s$ each. Hence, the spectrum
of regular $D3$ branes gives a moduli space which is a symmetric product
of the moduli space of a single regular brane. Namely
\begin{equation}
{\cal M}_N  = {\cal M}_1^N/ S_N
\end{equation} 

Notice that if we relax the invertibility of the matrices, then the
branes can only fractionate at the fixed complex lines, as we need
two out of the three matrices $x,y,z$ to be non-invertible in
order for the previous arguments to fail,  which is just 
as expected.

Consider now adding fractional $D3$ branes.
The configuration is allowed only if the gauge theory anomaly 
vanishes, that is, if the fractional D-brane is constructed by combinations of
effective $N=2$ directions \cite{LR}.
The effective $N=2$ directions are associated with 
the fixed complex lines of subgroups of $\Gamma$, on which
fractional branes can reside because the orbifold singularity is
not isolated. In principle, one can see 
a brane moving in the Coulomb branch only if it is a linear combination
of $N=2$ effective directions, and thus is automatically anomaly free.
This will be important when we compare with the supergravity analysis.

\section{Spectrum of chiral operators}\label{sec:spectrum}

In order to make the comparison with supergravity later on, we need to 
calculate the spectrum of single trace chiral primaries \cite{M,W,GKP}.

The quantum symmetry generators $Q_1$, and $Q_2$, which correspond
to the group of characters of $\Gamma$ act by
\begin{eqnarray}
Q_1&:&  \phi_{1,i,j} \to \alpha \phi_{1,i+1,j}\\
 &&\phi_{2,i,j} \to \alpha^{-1} \phi_{2,i+1,j}\\
 &&\phi_{3,i,j} \to \phi_{3,i+1,j}
\end{eqnarray}
\begin{eqnarray}
Q_2&:&  \phi_{1,i,j} \to  \phi_{1,i,j+1}\\
 &&\phi_{2,i,j} \to \beta \phi_{2,i,j+1}\\
 &&\phi_{3,i,j} \to \beta^{-1} \phi_{3,i,j+1}
\end{eqnarray}
and they only act by shifts on the gauge fields. It is obvious that they
are invariances of the field theory as long as all gauge couplings and
Yukawa couplings are equal, which is the orbifold point. This
identification of the quantum symmetry is necessary in order to define
the twist charge of operators in the field theory. In particular,
untwisted operators correspond to quantum symmetry singlets. The field
theory also has a $U(1)^3$ symmetry. One of the $U(1)$'s is the
$R$-symmetry and the other $U(1)$'s twist the field $\phi_1,
\phi_2,\phi_3$ by global phases.

Untwisted single-trace chiral primary operators are given schematically by
\begin{equation}
O_{abc} = \tr \phi_1^a\phi_2^b\phi_3^c
\end{equation}
for any closed loop in the quiver.  
In order for the operator to be a conformal primary, it must be such that
dragging one field around and using the cyclicity of the trace induces no
phase. The true operator is a linear combination of all of the possible terms
with the same structure that are related by the $F$ constraints.
The other linear combinations will correspond to chiral fields which are not 
primary. 

For a primary to be untwisted, it must satisfy a more stringent constraint. 
It is straightforward to show that untwisted primaries may be written as
\begin{equation}
\tr\ P\left(\phi_1^m,\phi_2^q,\phi_3^n,\phi_1\phi_2\phi_3,(\phi_1\phi_2)^n,
(\phi_2\phi_3)^m\right)
\end{equation}
where $P$ is any monomial. The point is that the variables in $P$ all
transform without phase under the quantum symmetry and commute with one
another, so that they are not constrained by the presence of discrete
torsion. These monomials are in one-to-one correspondence with monomials
in $x$, $y$, $z$, $u$, $v$, $w$ of the last section.  These operators
are in fact superconformal primaries, and additional conformal primaries
may be constructed by inserting factors of $W_\alpha^2$.

Twisted states come in families of the form 
\begin{equation}\label{eq:blotz}
\CO^{a}_1=\tr\ \phi_1^{am/s},
\end{equation}
for some integer $a$, in multiplicity $n/s$. Similar twisted operators
occur in $\phi_2$ and $\phi_3$ as well. These are operators where one
closes loops horizontally, vertically or diagonally. It is a non-trivial
fact that these operators exhaust the list of twisted superconformal
primaries. Note that the operator with $a$ a multiple of $s$ is
untwisted.

\section{Supergravity duals}\label{sec:ADS/CFT}

\subsection{Massless Twisted States and Monodromy}

The supergravity duals of the previous D-brane configurations are
obtained by looking at the near-horizon geometry of the $D3$-branes at
the orbifold, and they will correspond to geometries of the type
$AdS_5\times S^5/\Gamma$, where $\Gamma$ acts by isometries of the
$S^5$. Our first task is to understand how the orbifold encodes the
discrete torsion.

Because of our choice of group action, some  orbifold elements will
leave invariant three fixed circles of the $S_5$, which descend from the
fixed planes of the orbifold. Along these circles, we have locally a
singularity of the type $\CC^2/\ZZ_k$, where the $\ZZ_k$ is abelian and
generated by the element that fixes the point, that is, we will have  a
fibration of the $A_{k-1}$ singularities along these circles, which can
be locally resolved by $k-1$ blowup parameters into a bouquet of $k$
spheres which intersect according to the extended Dynkin diagram of the
$A_{k-1}$ group with one relation among these cocycles.

The twisted states are associated to elements of the group $\Gamma$.
They can survive in the supergravity limit if they correpond to massless
particles. It is known that zero modes for blowup modes are not present
in sufficient numbers to smooth out the geometry, so it is not possible
to study these models within classical geometry; we must turn to string
theory for detailed computations. Massless twisted states have support
at the fixed points of the group element they are associated with. Thus
at each fixed circle we have massless states in the string theory which
survive the low energy limit, and are associated with the group elements
that fix the circle.

By going around the fixed circle on a closed loop, we are actually
performing a twist by the group elements which don't fix the circle,
because the fixed circle is an invariant subspace of the other elements
of the orbifold group. The circle is of radius $k^{-1}$.

For the twisted strings that live at the orbifold circles, going around
the loop one picks up a phase equal to the discrete torsion of the cycle
acting on the group element to which the twisted state corresponds. This
is the discrete torsion phase in the partition function. In our case, it
sets the boundary conditions for the massless twisted sector states.

Consider for example the singularity of $z_2=z_3=0$ and the $g= e_2^\ell$
twisted sector corresponding to the subgroup which fixes the circle. 
When we set the boundary conditions for the  twisted field, we get
\begin{equation}\label{eq:bdcon}
\CO_g(t+2\pi h R/m) = \epsilon(g,e^h_1)\CO_g(t)
\end{equation}
for $t$ the geodesic coordinate along this circle measured in units of
$R$, the radius of the AdS space. This circle is of size $m^{-1}$, and
we wind around it $h$ times. Winding once around is the same as twisting
by $e_2$ in the unorbifolded covering space.

The blowup spheres are related to the discrete Fourier transform of the
twisted states, and the action of the discrete torsion phase (which one
can associate to the quantum symmetry of the local orbifold) becomes an 
automorphism of the Dynkin diagram which preserves the intersection form
and the orientation. Thus the geometry differs from the standard
$S^5/\Gamma$ space in that the singularities have monodromy of the
resolving spheres. This is a special example of monodromies that have
been studied in \cite{W2,BCD}, which are related to canonical
singularities \cite{R}.

The equation (\ref{eq:bdcon}) then becomes for the fields on the basis
of the spheres
\begin{equation}
\hat\CO_i(t+2\pi h R/m)
= \hat \CO_{i+a(m/s)h}(t)
\end{equation}
The precise discrete torsion value is encoded in the possible integer
values of $a$, which are relatively prime to $s$. Let us take $h=1=a$
for convenience. Then on going around the fixed circle the monodromy
acts as shown in Figure \ref{fig:monodromy}.

\FIGURE{\littlefig{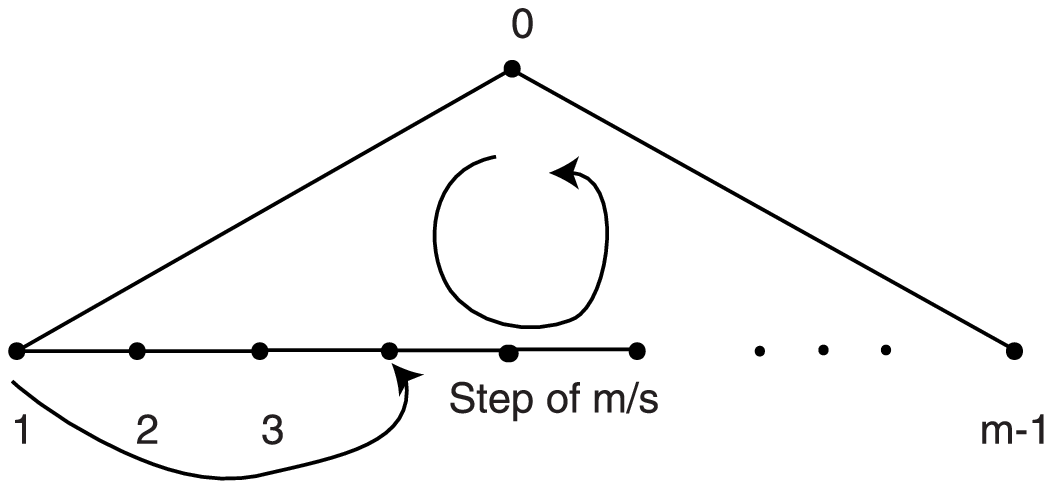}{2.5}\label{fig:monodromy}
\caption{Monodromy of spheres}}

It is worth pointing out that although we start with $k-1$ twisted
sectors, we have a bouquet of $k$ shrunken spheres. The relation in
homology makes one linear combination of the local fields untwisted, and
thus does not belong to the twisted string states. The counting of
twisted states is the same in both approaches.

Details of the exact calculation of the spectrum of masses are given in
the Appendix \ref{sec:sgspectrum}. The masses are given exactly by the
tree level mixing induced by the RR and NSNS backgrounds. It is not
difficult to see that the spectrum of masses of chiral twisted state
operators gives rise to conformal fields of the appropriate dimensions
and twist charge as the ones predicted by the orbifold low energy
effective field theory. In fact, one can see that the monodromy of the
spheres makes the fields periodic with period $2\pi R s/m$, as opposed
to $2\pi R/m$, so the circles look $s$ times larger than the geometric
circle, and the quantization of the masses of the states gives a factor
in the mass which is proportional to $m s^{-1} R^{-1}$ as opposed to the
factor of $m R^{-1}$ coming just from the size of the circle ({\it c.f.}
eq. (\ref{eq:blotz})). This is exactly the spectrum of states predicted by
the field theory too.

Notice also that the value of $a$ is irrelevant for this calculation, as
it amounts to a different permutation of the spheres when going around
the singularity. This is also in accord with the field theory. This is
the first check on the duality.

We should also point out that the spectrum of D-branes that can leave 
the singularity is also equal in both theories (supergravity and field theory),
 as any fractional D-brane
that is seen in the bulk of supergravity is either untwisted, or consists
of a collection of D-brane states that are fixed at the orbifold circles.

\subsection{Torsion Classes in Homology}

Notice that the singular manifold thus described has torsion on homology 
two-cycles. Upon monodromy on the circle,
different spheres are identified. 

On the singularity that we have just studied, there are $m/s$ equivalence classes
in homology at the singularity. They can be represented as the average
cycles
\begin{equation}
\tilde C_i = \frac1{s}[C_i+C_{i+m/s}+C_{i+2m/s}+\dots]
\end{equation}
and they intersect according to the $A_{m/s-1}$ Dynkin diagrm.

The relation in homology is given by 
\begin{equation}
\sum_i C_i = 0 = s\sum _i \tilde C_i
\end{equation}
so calling $T= \sum_i \tilde C_i$ we can prove that $s T=0$, but
$T\neq 0$ as a homology cycle. Indeed, if we wrap a brane around
the $T$ cycle, the brane is locally made of $m/s$ spheres, and is not 
allowed to leave the circle of singularities, since it can be made to 
correspond to a root of the $A_{m-1}$ lattice, and therefore it can 
not be deformed away from the singular cycle. This argument
works for $t$ copies of the cycle for all $t<s$. Thus the torsion
cycle is of order $s$.

This is seen also in the field theory, because one cannot solve the
$F$ constraints for $\phi_1, \phi_2,\phi_3$ all different from zero,
unless one has rank $s$ on each of the gauge groups in the quiver.

This establishes a direct correspondence between fractional regular
branes and the torsion classes of the manifold. The existence of a
torsion class is consistent with the absence of blowup modes in the
string spectrum. As the fractional $Dp$ branes also carry an anomalous
$D(p-2)$ brane charge \cite{GHM}(because the orbifold has constant B
field along each of the blowup cycles), one can have a BPS torsion
class with finite tension.

\section{New $N=1$ Dualities}\label{sec:Dualities}

Notice that the quivers we have obtained are identical to quivers of
standard type without discrete torsion. They only differ in the
superpotential, and in fact, they lie in the moduli space of
deformations of the standard quiver theories \cite{LS}. For a
single-node quiver, the field theory is a marginal deformation of the
$N=4$ theory
\begin{equation}
W=\sqrt{2}g\tr \phi_1 [\phi_2,\phi_3]-
\sqrt{2}g(\lambda^s-1)\tr \phi_1\phi_2\phi_3
\end{equation}
Alternatively, this may be rewritten in terms of $f_{abc}$ and $d_{abc}$
as appears in Ref. \cite{LS}. In the case of multinode quiver theories,
the superpotential is a marginal deformation of standard $N=1$ quiver
theories.

Thus, with the AdS/CFT dictionary in hand, these field theory
deformations act by geometrically deforming the theory which started
with $S^5$, or $S^5/\Gamma$. In particular, we expect that the marginal
deformation corresponds to replacing the 5-sphere with a deformed
version, and also turning on 2-form fields.

In supergravity it has been argued that one cannot observe some of these
deformations \cite{GPPZ}. However, as we deform the superpotential, the
conditions for a field being primary change, as the moduli space of
couplings is curved. In the field theory this is seen because conformal
invariance imposes restrictions on the couplings.\cite{LS}

The moduli space of the theory changes when we follow
these deformations. In particular, we have to find solutions of the
$F$-constraints, which take the form
\begin{equation}
\phi_1\phi_2 = \omega \phi_2\phi_1
\end{equation}
For generic $\omega$, there is no solution to these matrix equations
with $\phi_1\phi_2\neq 0$ at finite $N$, and therefore the classical
moduli space disappears. But on the other hand, we can elevate the
previous equation to non-commutative geometry as in \cite{CDS}, if we
let $\phi_i$ be operators acting on an infinite dimensional Hilbert
space. In this case we would be forced onto the large $N$ field theory
from the start.

As a marginal perturbation, $\lambda$ would seem to be a continuous
parameter. However, properties of the theory change discontinuously as
we change $\lambda$. There is a direct analogy here to the
$\theta$-angle of six-dimensional gauge theories,\cite{W2} which are
also related to non-commutative geometry.\cite{LR2} For values of
$\lambda$ which are roots of unity, the classical moduli space has
solutions for finite matrices, and it is exactly in this case where we
have the dual theory with the orbifold.

Thus, for each of these theories under the conditions for the Maldacena
conjecture to be reduced to supergravity, we have two types of
supergravity duals. Namely, $AdS_5\times \tilde S_5$  where we have a
deformed sphere $\tilde S^5$ and $AdS_5\times S^5/\Gamma_d$, where we
use the subscript $d$ to indicate the presence of discrete torsion.
(This is the extreme case where the quiver theory has only one node, but
one can restate it for the general case by writing the extra quotients
properly.) These theories are mirror to one another, even at finite $N$.

At infinite $N$, there are additional field theory dualities relating
different values of $\lambda$. On $AdS_5\times S^5/\Gamma_d$ we have
various choices of discrete torsion parameter. But it can easily be seen
that neither the chiral ring nor the classical moduli space depends on
this choice, once $s$ is fixed. Moreover one sees that the discrete
torsion parameter can only be measured in chiral correlation functions
involving twist fields that have support at different singularities.
In the supergravity approximation, in order for twist fields
to interact, they should be able to be at the same spacetime point. Thus
all of these amplitudes are suppressed because the twisted strings have
to extend far away along the sphere in order to interact.

On the field theory side, these amplitudes vanish in the free field
limit, as they would involve, for example, the OPE of $\tr(\phi_1^k)
\tr(\phi_2^k) \tr(\phi_3^k)$. Notice that if the free field limit is the
correct way to calculate three point functions (just as in $AdS_5\times
S^5$)\cite{LMRS} then the discrete torsion parameter disappears in the
supergravity regime. Indeed, discrete torsion appears only in a non-planar 
diagram, as in Figure \ref{fig:hexagon}.

It would be natural then to identify all of these $CFT$ theories as dual
pairs. They meet all of the requirements of $N=1$ duality, as they have
the same moduli space of flat directions and the same spectrum of chiral
primaries. At large $N$ they also have the same set of three point
correlation functions.

This duality would act by exchanging the fundamental roots of unity. 
It is not continous on the parameter $\lambda$, but it is a well
known mathematical object: it is the Galois group of the algebraic
extension $\QQ(\tors)$. In our case this Galois group can also be made
to act on the algebra generated by the matrices $x,y,z,u,v,w$ if all of
their components are algebraic numbers, but the significance of this
curiosity is at best obscure.

\section{Conclusions}

We have analyzed some Abelian orbifolds with discrete torsion from the
D-brane perspective. Our results show that the simple cases give rise to
deformations of known theories of the brane box type, which differ only
in the superpotential. This result extends the work of \cite{D,DF}.
We tested the field theory by reproducing the orbifold as the moduli
space of D-branes and by calculating the spectrum of chiral primaries.

In analyzing the field theory with the AdS/CFT correspondence, we
discovered a new geometric picture for the orbifolds with discrete
torsion. Indeed we found that as the fixed planes of singularities of
the orbifold become fixed circles in the AdS/CFT correpondence, the
study of massless twisted states on these fixed circles revealed that
the blowup modes of the singularities have monodromy.  The resolved
geometry then has a different topology than the standard orbifold, and
provides a geometrical understanding of the discrete torsion. Also, the
obstructions to resolving the Calabi-Yau singularities may be traced
back to the existsence of monodromy of the blowup modes. We have
explicitly checked this picture by calculating the spectrum of chiral
primaries in the supergravity/string theory and comparing with the field
theory.

The AdS/CFT correspondence can then be used to prove new examples of
duality. Indeed, as the theories with discrete torsion are seen to be
given by deformations of standard orbifolds, one can follow these
deformations at the supergravity level. This provides new examples of
mirror symmetry, as both string theory descriptions are weakly coupled.

In the large $N$ limit we have also found new dualities. These dualities
appear because the spectrum of operators in the supergravity is
insensitive to the discrete torsion parameter, and the three-point
functions of chiral twisted states that could measure this parameter
vanish in this limit. The duality acts discontinuosly on the parameters
defining the theory by permuting the different roots of unity.

It would be interesting to get a better understanding of the deformation
theory and resolution of singularities within the D-brane language along
the lines of \cite{BGLP}, and to extend these results for other marginal
and relevant deformations of the field theory. We would also like to
extend the discrete torsion models to other geometries, wuch as the
conifold\cite{KW,MP,U}. This is work in progress \cite{BL3}.

\acknowledgments Partially supported by the United States
Department of Energy, grant DE-FG02-91ER40677 and 
an Outstanding Junior Investigator Award. 

\appendix
\section{Supergravity spectrum of chiral operators}\label{sec:sgspectrum}

In this appendix we set up the calculation of the supergravity spectrum
of states in the orbifolds with discrete torsion. Our construction is
done following the background normal coordinate expansion for
Ramond-Ramond background type of calculations explained in \cite{BL,
BL2}. These results coincide with supergravity calculations obtained in
\cite{G} in cases where smooth geometry exists, but generalizes to
spaces which have less local symmetry. For simplicity we will study the
local $\CC^2/\ZZ_2$ fixed circles (only because it is easier to write
the twisted vertex operators). The results generalize immediately to any
other orbifolds.

The analysis is completely standard. The spin field $S^\alpha$ is
written in terms of $Spin(5,1)\times Spin(4)$ spin-fields: $S^\alpha\to
\{S^A\Sigma^a,S_A\Sigma^{\dot a}\}$. Conventions for operator products
may be found in the Appendix \ref{sec:OPE}.

\subsection{Twisted States of the $\CC^2/\ZZ_2$ Orbifold}

At the fixed circle, we have a $\CC^2/\ZZ_2$ orbifold. The twisted sector
massless strings are given by the following operators.

In the $NSNS$ sector, we find
\begin{equation}\label{eq:NSop}
\CO_{NS}^{ab}\sim \Sigma^a\tilde\Sigma^b e^{-\phi}e^{-\tilde\phi}\twfld
\end{equation}
These transform as $({\bf 1,(1+3,1)})$ under $Spin(5,1)\times Spin(4)$.
and correspond to the $NS$ $\theta$-angle, $\Theta_{NS}$ and the triplet
of blowup modes.

In the $RR$ sector, we find
\begin{equation}\label{eq:Rop}
\CO_{R,AB}\sim S_A\tilde S_B e^{-\phi/2}e^{-\tilde\phi/2}\twfld
\end{equation}
which transform as $({\bf 6+10,(1,1)})$. The ${\bf 6}$ is the field
strength of the $RR$ $\theta$-angle, $\pa_\mu\Theta_{R}$.

\subsection{The Background}\label{sec:Background}
\newcommand{\fud}{\alpha}

The background that we are interested in must be a singlet  under
both $Spin(4)$ and $Sp(2)\sim Spin(5)\subset Spin(5,1)$. The $RR$ part is then
determined as follows. The ${\bf 6}$ of $Spin(5,1)$ has one $Sp(2)$
singlet, while the ${\bf 10}$ has none. We thus find that the $RR$
background vertex is
\begin{equation}
V_{RR}\sim  h\int\left( J_{AB}\epsilon_{ab} S^A\Sigma^a\tilde S^B
\tilde\Sigma^b +\fud J^{AB}\epsilon_{\dot a\dot b} S_A\Sigma^{\dot a}
\tilde S_B\tilde\Sigma^{\dot b}\right)e^{-\phi/2}e^{-\tilde\phi/2}
\end{equation}
where $h$ is a normalization constant, $\fud$ is a fixed phase
and $J$ is the antisymmetric $Sp(2)$ invariant.

First, let us note that two $RR$ background insertions give a log
divergence when they are close together
\begin{equation}
V_{RR}(z)V_{RR}(z')\sim 16 h^2 \ln\varepsilon\int\left(\psi^\mu\tilde\psi_\mu
-\psi^M\tilde\psi_M\right)e^{-\phi}e^{-\tilde\phi}
\end{equation}
The signature here is $(5,5)$, and this divergence corresponds to a
contribution to the graviton $\beta$-function. The other contribution
may be computed directly within the normal coordinate expansion, and
takes the form
\begin{equation}
\frac14\log\varepsilon \int R_{ij}\psi^i\tilde\psi^je^{-\phi}e^{-\tilde\phi}
\end{equation}

Thus we fix the normalization of the $RR$ background
\begin{equation}
hR=\frac{1}{ 2\sqrt{2}}
\end{equation}
where $R$ is the radius of the $AdS_5$ metric.

\subsection{Mixing and Masses}

Here we compute the masses of twisted eigenstates. In the presence of
the $RR$ background, there is operator mixing at lowest order. In 
particular, we find mixing between $\Theta_{NS}$ and $\Theta_{R}$, as
follows. Using the notations (\ref{eq:NSop}),(\ref{eq:Rop}), 
these singlets have vertex operators of the form
\begin{eqnarray}
\CO_{NS}^{(1)}&\sim&\Theta_{NS}
\epsilon_{ab}\CO_{NS}^{ab} 
\\
\CO_{R}^{(1)}&\sim&\pa_\mu\Theta_{R}(\Gamma^\mu)^{AB}\CO_{R,AB}
\end{eqnarray}
Mixing is induced through the Fischler-Susskind mechanism -- there are
logarithmic divergences  between the background $V_{RR}$ and these
operators. 

Consider the equation of motion for the $RR$ singlet. We find mixing 
through
\begin{equation}
\CO_{NS}^{(1)}(z)\cdot V_{RR}(z')\sim 2h\Theta_{NS}\
J_{AB}\ \ksl^{AC}\ksl^{BD} \log\varepsilon\ \CO_{R,CD}(z').
\end{equation}
Thus we find an equation of motion
\begin{equation}
\frac{1}{ 2}k^2 \pa_\mu\Theta_{R}(\Gamma^\mu)^{CD}
+2hJ_{AB}\ksl^{AC}\ksl^{BD}\Theta_{NS}=0
\end{equation}
and thus mass terms
\begin{equation}
\frac {n^2}{ R^2}\Theta_R+4ih{\frac{n}{ R}}\Theta_{NS}
\end{equation}
Similarly, mixing for the $NSNS$ singlet is induced through
\begin{equation}
\CO_{R}^{(1)}(z)\cdot V_{RR}(z')\sim h\pa_\mu\Theta_R\
J_{AB}(\Gamma^\mu)^{AB}\ \log\varepsilon\ 
\epsilon_{ab}\CO_{NS}^{ab} (z')
\end{equation}

Putting these results together, we have a mass matrix
\begin{equation}\label{eq:masses}
[m^2]=\begin{pmatrix} (\frac{k}{ R})^2 & -8ih (\frac{k}{ R})\\
+4ih(\frac{k}{ R}) & (\frac{k}{ R})^2
\end{pmatrix}
\end{equation}
where $k$ is the momentum along the fixed circle in units of $R$. 

Within the AdS-CFT correspondence\cite{M,W,GKP}, we expect that
the eigenvalue  masses correspond to the dimensions of operators in the 
boundary CFT:
\begin{equation}
\Delta=2+\sqrt{4+m^2R^2}
\end{equation}
Given the result \ref{eq:masses}, we find a tower of states
\begin{equation}
\Delta_1=2+|k\pm 2|
\end{equation}
where $k\in\ZZ$. 

Similiarly, one obtains for the triplets
\begin{equation}
\Delta_3=2+|k|
\end{equation}
The interesting part of this calculation comes from the origin of the
mass shift, which can be traced to the $\sigma$-model coupling
proportional to $R_{abcd}\psi^a\psi^b\bar\psi^c\bar\psi^d$, and thus for
spaces where $R$ is not symmetric one should be able to observe the
effect here.

It is clear that these calculations are independent of the fact 
that we had a $\CC^2/\ZZ_2$ orbifold. The structure of states
is the same for any orbifold group.
The different orbifolds $\CC^2/\ZZ_n$ will have instead $n-1$ twisted
sectors. 

The singlet states are chiral.
Because of the monodromy of the circle $k =a m/s$, and the dimensions
of the operators match the $\tr(\phi_i^{(am/s)})$, with multiplicities. 
The other operators are related to $\tr(W^2\phi_i^{(am/s)})$.

\section{Conventions for spin fields}  \label{sec:OPE}

Under $Spin(5,1)\times Spin(4)\subset Spin(9,1)$, the spin fields
decompose as $S^\alpha\to \{S^A\Sigma^a,S_A\Sigma^{\dot a}\}$. Here $A$
is an $Spin(5,1)$ spinor index and $a,\dot a$ are 
$Spin(4)=SU(2)\times SU(2)$
indices. We denote vector indices by $\mu,\nu,\ldots$ for $Spin(5,1)$ and
$M,N,\ldots$ for $Spin(4)$. 

Operator products are
\begin{eqnarray}
\Sigma^a(z)\Sigma^b(0)\sim \frac{\epsilon^{ab}}{ z^{1/2}}  &&\\
\Sigma^{\dot a}(z)\Sigma^{\dot b}(0)\sim\frac {\epsilon^{\dot a\dot b}
}{ z^{1/2}} &&\\
\Sigma^a(z)\Sigma^{\dot a}(0)\sim (\Gamma_M)^{a\dot a}\psi^M(0)  &&\\ 
\Sigma^{\dot a}(z) \Sigma^a(0)\sim (\Gamma_M)^{\dot a a}\psi^M(0)&&
\end{eqnarray}

\begin{eqnarray}
\psi^M(z)\Sigma^a(0)\sim\Sigma^a(z)\psi^M(0)\sim  \frac{1}{ z^{1/2}}
(\Gamma^M)^{a\dot a}\Sigma_{\dot a}(0)&&\\
\psi^M(z)\Sigma^{\dot a}(0)\sim\Sigma^{\dot a}(z)\psi^M(0)
\sim  \frac{1}{ z^{1/2}}
(\Gamma^M)^{\dot a a}\Sigma_{a}(0)&&
\end{eqnarray}

The matrices $\Gamma^M$ have Clifford normalization. Similarly
\begin{equation}
S^A(z)S_B(0)\sim S_A(z)S^B(0)\sim\frac{\delta^A_B}{ z^{3/4}}
\end{equation}
\begin{equation}
S^A(z)S^B(0)\sim \frac{1}{ z^{1/4}}(\Gamma_\mu)^{AB}\psi^\mu(0)
\end{equation}
\begin{equation}
S_A(z)S_B(0)\sim  \frac{1}{ z^{1/4}}(\Gamma_\mu)_{AB}\psi^\mu(0)
\end{equation}
\begin{equation}
\psi^\mu(z)S^A(0)\sim \frac{1}{ z^{1/2}}(\Gamma^\mu)^{AB}S_{B}(0)
\end{equation}
\begin{equation}
\psi^\mu(z)S_A(0)\sim \frac{1}{ z^{1/2}}(\Gamma^\mu)_{AB}S^{B}(0)
\end{equation}
\begin{equation}
S^A(z)\psi^\mu(0)\sim {\frac{1}{ z^{1/2}}}S_{B}(0)(\Gamma^\mu)^{BA}
\end{equation}
\begin{equation}
S_A(z)\psi^\mu(0)\sim {\frac{1}{ z^{1/2}}}S^{B}(0)(\Gamma^\mu)_{BA}
\end{equation}
where $(\Gamma_\mu)_{AB}=J_{AC}(\Gamma_\mu)^{CD}J_{DB}$, with $J$ the
antisymmetric $Sp(2)$ invariant.

%
\providecommand{\href}[2]{#2}\begingroup\raggedright\endgroup

\end{document}